\documentclass[a4paper,10pt]{article}
\usepackage[dvips]{graphicx}
\usepackage{amssymb,amsmath}
\numberwithin{equation}{section}

\begin{document}

\title{ A note on the relationship between solutions of Einstein, Ramanujan and Chazy  equations}
\author{Kuralay Esmakhanova$^1$, Yerlan Myrzakulov$^1$,  Gulgasyl Nugmanova$^1$ \\and  Ratbay Myrzakulov$^{1,2}$\footnote{The corresponding author. Email: rmyrzakulov@csufresno.edu; rmyrzakulov@gmail.com}\\ \textit{$^1$Eurasian International Center for Theoretical Physics, } \\ \textit{Eurasian National University, Astana 010008, Kazakhstan} \\ \textit{$^2$Department of Physics, CSU Fresno, Fresno, CA 93740 USA}}

%\date{}

\maketitle
\begin{abstract} The Einstein equation for the  Friedmann-Robertson-Walker metric  plays a fundamental  role in cosmology. The direct search of the exact  solutions of the Einstein equation even in this simple metric case is sometime a hard job. Therefore, it is useful to construct solutions of the Einstein equation using a known solutions of some other equations which are equivalent or related  to the Einstein equation. In this work, we establish the relationship   the  Einstein equation with two other famous equations namely  the Ramanujan equation  and the Chazy equation.  Both these two equations play an imporatant role in the number theory. Using the known solutions of the Ramanujan and Chazy equations, we find the corresponding solutions of the Einstein equation. The relationship between the Friedmann equation and the equations of the trefoil knot and figure-eight knot is established.
\end{abstract}
\vspace{2cm} 

\sloppy

%\tableofcontents
\section{Introduction} In this paper we investigate the Einstein equation for the Friedmann-Robertson-Walker (FRW) metric. The direct search of the exact  solutions of the Einstein equation even in this simple metric case is sometime a hard job. Therefore, it is useful to construct solutions of the Einstein equation using a known solutions of some other equations which are equivalent or related to the Einstein equation. One of known differential equation from the number theory is  the Ramanujan equation (RE). It is well studied. In particular, it is known that the RE is equivalent to the famous Chazy-III equation (CE-III). In the literature, many exact solutions of the RE and the CE-III are presented.  Both these two equations play an imporatant role in the number theory.  
 The goal of this work is to establish the relationship between three equations namely the Einstein equation, the RE  and the CE-III. 
\section{The Einstein equation}
In this section we briefly recall some basic facts about the Einstein  equation for the FRW metric. 
We start from the standart gravitational action
\begin {equation}
S=\int d^{4}x\sqrt{-g}[R+L_m],
\end{equation} 
 where $R$ is the scalar curvature,  $L_m$ is the Lagrangian for the matter. Consider the FRW metric
	\begin{equation}
ds^2=dt^2-a^2(dx^2+dy^2+dz^2).
\end{equation}
Then the Einstein equation reads as
	\begin{eqnarray}
	3H^2-\rho&=&0,\\ 
		2\dot{H}+3H^2+p&=&0,\\
			\dot{\rho}+3H(\rho+p)&=&0.
	\end{eqnarray} Let us rewrite the equation (2.4) as
  	\begin{equation}
\dot{F}=\frac{F^2-E}{12},
\end{equation}
where
	\begin{equation}
F=-18H,\quad E=-108p=108(2\dot{H}+3H^2).
\end{equation}
Now we introduce the new function $J$ as
	\begin{equation}
J=FE-3\dot{E}=-2\cdot 18^2(\ddot{H}+9H\dot{H}+9H^3)
\end{equation}
and  demand that this function satisfies the equation: $2\dot{J}=FJ- E^2$. So finally we come to the following system:
\begin{eqnarray}
	\dot{F}&=&\frac{F^2-E}{12},\\ 
		\dot{E}&=&\frac{FE-J}{3},\\
			\dot{J}&=&\frac{FJ- E^2}{2}.
	\end{eqnarray} 			
It is the famous Ramanujan  equation (RE) \cite{Ramanujan1}-\cite{Ramanujan2}.	
We can rewrite it as
\begin{eqnarray}
	qF_q&=&\frac{F^2-E}{12},\\ 
		qE_q&=&\frac{FE-J}{3},\\
			qJ_q&=&\frac{FJ- E^2}{2},
	\end{eqnarray} 
	where $q=e^{t}$. Then the system (2.12)-(2.14) has the following solution  (see e.g. \cite{Chakravarty2})
	\begin{eqnarray}
	F&=&1-24\sum_{n=1}^{\infty}\frac{nq^n}{1-q^n},\\ 
		E&=&1+240\sum_{n=1}^{\infty}\frac{n^3q^n}{1-q^n},\\
			J&=&1-504\sum_{n=1}^{\infty}\frac{n^5q^n}{1-q^n},
	\end{eqnarray}
which are in fact Ramanujan's Eisenstein series. 	If we eliminate $E$ and $J$ from the system (2.9)-(2.11), we deduce the following differential equation for $F$:
		\begin{equation}
\dddot{F}=F\ddot{F}-1.5\dot{F}^{2}. 
\end{equation}
	This equation after the scale transformation 
		\begin{equation}
F=2y
\end{equation}
	takes the form
		\begin{equation}
\dddot{y}=2y\ddot{y}-3\dot{y}^{2}.
\end{equation}
It is the Chazy-III  equation (CE-III) \cite{Chazy1}. In recent years, it has been shown that the CE-III (2.20) arises in several areas of mathematical physics including magnetic monopoles \cite{Atiyah}, self-dual Yang-Mills and Einstein equations \cite{Chakravarty1}-\cite{Hitchin2}, and topological field theory \cite{Dubrovin}. In addition, the CE-III has been derived as special reductions of hydrodynamic type equations \cite{Ferapontov} as well as stationary, incompressible  Prandtl boundary layer equations \cite{Rosenhead}. Recently, the geometry of some Chazy equations were studied in \cite{Guillot}.  We remark that the CE-III  (2.20) is equivalent to the following  equation 
	\begin{eqnarray}
	\dot{A}&=&BC-A(B+C),\\ 
		\dot{B}&=&AC-B(A+C),\\
			\dot{C}&=&AB-C(A+B)
	\end{eqnarray} 	
	if one sets
		\begin{eqnarray}
	y&=&-2(A+B+C),\\ 
		\dot{y}&=&2(AC+BC+AB),\\
			\ddot{y}&=&-12ABC.
	\end{eqnarray}
It is the classical Darboux-Halphen equation \cite{Ablowitz2} which appeared in Darboux's analysis of triply orthogonal surfaces and later solved by Halphen. The Darboux-Halphen equation (2.21)-(2.23)   is also equivalent to the vacuum Einstein equations for Riemannian self-dual Bianchi-IX metrics \cite{Gibbons}-\cite{Hitchin1}.	
\section{Solutions}	
As is known, the CE-III  has some exact solutions (see e.g.  \cite{Chakravarty2}, \cite{Ablowitz2}, \cite{Ablowitz1}, \cite{Sasano}, \cite{Brezhnev}). For example, it has the following    solutions:
		\begin{equation}
y_1=c_1=const, \quad y_2=-\frac{6}{t}, \quad y_3=\frac{\alpha}{t^2}-\frac{6}{t}, \quad y_4=4\frac{d}{dt}\log{\theta_{1}^{'}(0,t)}, \quad \alpha=const
\end{equation}
	and
		\begin{equation}
y_5(t)=\pi iF(e^{2\pi it}), \quad Im(t)>0.
\end{equation}In this section, we construct the solutions of the Einstein equation (2.3)-(2.5) corresponding to the solutions $y_1, y_2, y_3$ of the CE-III (2.20) that same to the RE (2.9)-(2.11).
\subsection{Example 1.}
First of all, let us consider the solution $y_2$ (3.1). The corresponding solution of the RE 
(2.9)-(2.11) is given by
	\begin{equation}
F_2=-\frac{12}{t},\quad E_2=0, \quad J_2=0.
\end{equation}
For the solution $y_2$ we get the following solution of the Einstein equation
\begin{eqnarray}
a&=&a_0t^{\frac{2}{3}},\\ 
	H&=&\frac{2}{3t},\\ 
		\rho&=&\frac{4}{3t^2}.\\
			p&=&0.
	\end{eqnarray} 	
Hence we have
	\begin{equation}
\rho+p=\frac{4}{3t^2}.
\end{equation}
Also we can directly check that the continuity equation (2.5) 
satisfies identically. The equation of state parameter is $w=0$.
\subsection{Example 2.}
Now we consider the case when $y=y_1=c_1=-9H_0=const$. Then  the RE 
(2.9)-(2.11) has the following solution
	\begin{equation}
F_1=-18H_0,\quad E_2=324H_0^2, \quad J_2=-5832H_0^3.
\end{equation}
For the solution $y_1$ we get
\begin{eqnarray}
a&=&a_0e^{H_0t},\\ 
	H&=&H_0,\\ 
		\rho&=&3H_0^2,\\
			p&=&-3H_0^2.
	\end{eqnarray} 	
In this case,  the continuity equation (2.5) 
satisfies identically and the equation of state parameter is $w=-1$.
\subsection{Example 3.}
Now we consider the solution $y_3$ (3.1). The corresponding solution of the RE 
(2.9)-(2.11) is
	\begin{equation}
F_3=2(\frac{\alpha}{t^2}-\frac{6}{t}),\quad E_3=\frac{4\alpha^2}{t^4}, \quad J_3=\frac{8\alpha^3}{t^6}.
\end{equation}
In this case, the scale factor is
	\begin{equation}
a=a_0t^{\frac{2}{3}}e^{\frac{\alpha}{9t}}.
\end{equation}
For the Einstein equation we have
\begin{eqnarray}
	H&=&-\frac{1}{9}(\frac{\alpha}{t^2}-\frac{6}{t}),\\ 
		\rho&=&\frac{1}{27}(\frac{\alpha}{t^2}-\frac{6}{t})^2,\\
			p&=&-\frac{\alpha^2}{27t^4}.
	\end{eqnarray} 	
Using 
	\begin{equation}
\rho+p=\frac{4}{3t^2}-\frac{4\alpha}{9t^3},
\end{equation}
we can directly check that the continuity equation (2.5) 
satisfies identically. The equation of state parameter is 	\begin{equation}
w=-\frac{\alpha^2}{(\alpha-6t)^2}.
\end{equation}
\section{Conclusion} 
In summary, the Einstein equation in FRW metric is studied in this work. The relationship between the Einstein, Ramanujan and Chazy-III equations is established. Using this result that is using some known solutions of the RE and CE-III, some exact solutions of the Einstein equation are constructed.  Finally, we note that it is interesting to find the relationship between the Einstein equation with the other equations known from the other branches of physics and/or mathematics. Consider some examples (below we assume that all variables are dimensionless).  

i) \textit{The trefoil knot universe}. Let the EoS has  the following parametric form $	\rho=3r^2\cos^2 2t, 	p=6\sin 3t\cos 2t+4r\sin 2t-3r^2\cos^2 2t, $
	where $r=2+\cos 3t$. Then we have $H=r\cos 2t.$
Let us now we introduce three new variables as $x=r\cos 2t, 	y=r\sin 2t, 			z=\sin 3t.$ 
It is nothing but the parametric equations for  the trefoil knot. Hence this example we call the  trefoil knot universe or the  trefoil  knot universe model.

ii)	\textit{The figure-eight knot universe}. Let us now we consider the   EoS of the form $	\rho=3h^2\cos^2 3t, 		p=4\sin 2t\cos 3t+6h\sin 3t-3h^2\cos^2 3t, $
	where $h=2+\cos 2t$. 
	Then $H=h\cos 3t.$ 
As above, let us we introduce three new functions as: $	x=h\cos 3t, 		y=h\sin 3t, 			z=\sin 4t.$ It is  the parametric equations of 	the figure-eight knot.  In this sense this example  we call the  figure-eight knot universe or the  figure-eight knot universe model.

iii) \textit{The Lorenz oscillator equation}.  Let us rewrite the equation (2.4) as
	\begin{equation}
X_N=\sigma(Y-X),
\end{equation}
where 
	\begin{equation}
X=H^2,\quad Y=\frac{(\sigma-3)X-p}{\sigma}, \quad N=\ln{a}, \quad \sigma=const.
\end{equation}
Now we introduce a new function $Z$ as
	\begin{equation}
Z=\frac{\delta X-Y-Y_N}{X},
\end{equation}
where $\delta=const$. Then these three functions $X, Y, Z$ satisfy the equations
\begin{eqnarray}
X_N&=&\sigma(Y-X),\\ 
	Y_N&=&X(\delta-Z)-Y.
	\end{eqnarray}
Our next step is the following constraint for $Z$ namely it must obeys the additional equation $Z_N=XY-\beta Z$. So finally we get the system
\begin{eqnarray}
X_N&=&\sigma(Y-X),\\ 
	Y_N&=&X(\delta-Z)-Y,\\
Z_N&=&XY-\beta Z.
	\end{eqnarray}
It is the equations of Lorenz oscillator. So as the RE and the CE-III, the Lorenz oscillator equation is modelled some subclass of  solutions of the Einstein equation. We note that the Lorenz equation (4.6)-(4.8) can be rewritten as the third order differential equation for X(t) \cite{Sen}-\cite{MR444}:
	\begin{equation}
X\dddot{X}-\dot{X}\ddot{X}+X^3\dot{X}+\sigma X^4+(\beta+\sigma+1)X\ddot{X}+(\sigma+1)(\beta X\dot{X}-
\dot{X}^2)+\beta(1-\delta)\sigma
X^2=0.
\end{equation}
   
  \section{Appendix: A list of Chazy equations}
  In \cite{Chazy1}, Chazy attempted the complete classification of all  third-order differential equations of the form:
  	\begin{equation}
y^{'''}=F(t,y,y^{'},y^{''}),
\end{equation}
where $F$ is a polynomial in $y, y^{'}$ and $y^{''}$ and locally analytic in $z$, having the Painleve property, $y^{'}=dy/dz$ and so on.. Here we present a list of the canonical reduced Chazy equations  (see e.g. \cite{Sasano}) : 
  \begin{eqnarray}
CE-I: y^{'''}&=&-6y^{'2},\\ 
	CE-II: y^{'''}&=&-2yy^{''}-2y^{'2},\\
	CE-III: y^{'''}&=&2yy^{''}-3y^{'2},\\
	CE-IV: y^{'''}&=&-3yy^{''}-3y^{'2}-3y^2y^{'},\\
	CE-V: y^{'''}&=&-2yy^{''}-4y^{'2}-2y^2y^{'},\\
	CE-VI: y^{'''}&=&-yy^{''}-5y^{'2}-y^2y^{'},\\
	CE-VII: y^{'''}&=&-yy^{''}-2y^{'2}+2y^2y^{'},\\
	CE-VIII: y^{'''}&=&6y^2y^{'},\\
	CE-IX: y^{'''}&=&12y^{'2}+72y^2y^{'}+54y^4,\\
	CE-Xa: y^{'''}&=&6y^2y^{'}+\frac{3}{11}(9+7\sqrt{3})(y^{'}+y^2)^2,\\
	CE-Xb: y^{'''}&=&6y^2y^{'}+\frac{3}{11}(9-7\sqrt{3})(y^{'}+y^2)^2,\\
	CE-XI: y^{'''}&=&-2yy^{''}-2y^{'2}+\frac{24}{n^2-1}(y^{'}+y^2)^2,\\
	CE-XII: y^{'''}&=&2yy^{''}-3y^{'2}-\frac{24}{n^2-36}(6y^{'}-y^2)^2,\\
	CE-XIII: y^{'''}&=&12yy^{'}.
	\end{eqnarray} 	

	\end{document}